\documentclass[aps,prl,twocolumn]{revtex4}

\usepackage{amsmath}
\usepackage{amssymb}
\usepackage{graphicx}

\begin{document}

\title{Comment on "Molecular Transport Junctions: Clearing Mists"}

\author{Massimiliano Di Ventra}
\affiliation{University of California - San Diego, La Jolla, California 92093}

\maketitle In a recent review article~\cite{LindRat} Lindsay and
Ratner wrongly report several results found in literature regarding
the calculation of the linear-response conductance of a
1,4-phenyldithiol molecule connected to gold electrodes. This leads
them to the wrong conclusion regarding the agreement between the
different theoretical calculations for this system. In table 2 of
their paper, they report that the conductance of this molecule
calculated with static density-functional theory (DFT) with metal
electrodes described using the jellium model is 1 $\mu S$. This is
not correct. The linear-response conductance of that model is 3 $\mu
S$ as calculated in [2]. In addition, the conductance of reference
[15] in their paper, whose method they call "DFT+bulk states" and
which they use as the "reference theory", is about 5 $\mu S$ and not
47 $\mu S$ as stated in [1]. This value is very close to the one
obtained using the jellium electrodes [2,3]. To the short list in
table 2 of Ref. [1], one can add several other results, obtained
using static DFT with various computational details. Here I just
mention the conductance calculated by Ratner and co-workers, where
in Ref. [4] they obtain a value of 2.8 $\mu S$ and in Ref. [5] they
report 4.8 $\mu S$. In Ref. [6], a conductance of 6 $\mu S$ is
calculated by Damle {\em et al}. These results show that the
statement made in Ref. [1] "With the exception of the results of the
jellium model, the other values [of conductance] are within an order
of magnitude" of those reported in Ref. [15] of [1], is incorrect.
In fact, the opposite seems to hold: the majority of calculations
yields a linear-response conductance close to the one obtained using
the jellium model [2]. These values, however, are still too large
compared to the experimental ones for the same system.

\end{document}